\documentclass[5p]{elsarticle}

\biboptions{sort&compress}

\usepackage{url}
\usepackage{amsmath}
\journal{Reports of Practical Oncology and Radiotherapy}

\title{Reducing the dosimetric impact of positional errors in field
  junctions for craniospinal irradiation using VMAT}

\author{Andrej Strojnik}
\ead{astrojnik@onko-i.si}
\author{Ignasi M\'endez}
\ead{nmendez@onko-i.si}
\author{Primo\v{z} Peterlin\corref{cor}}
\ead{ppeterlin@onko-i.si}

\cortext[cor]{Corresponding author. \ead{ppeterlin@onko-i.si}}
\address{Institute of Oncology Ljubljana, Zalo\v{s}ka 2, SI-1000
  Ljubljana, Slovenia}

\begin{document}
\begin{abstract}
  \paragraph{Aim}
  To improve treatment plan robustness with respect to small shifts in
  patient position during the VMAT treatment by ensuring a linear
  ramp-like dose profile in treatment field overlap regions.

  \paragraph{Background} 
  Craniospinal irradiation (CSI) is considered technically challenging
  because the target size exceeds the maximal field size, which
  necessitates using abutted or overlapping treatment fields.
  Volumetric modulated arc therapy (VMAT) is increasingly being
  examined for CSI, as it offers both better dose homogeneity and
  better dose conformance while also offering a possibility to create
  field junctions which are more robust towards small shifts in
  patient position during the treatment.

  \paragraph{Materials and Methods} 
  A VMAT treatment plan with three isocenters was made for a test case
  patient.  Three groups of overlapping arc field pairs were used; one
  for the cranial and two for the spinal part. In order to assure a
  ramp-like dose profile in the field overlap region, the upper spinal
  part was optimised first, with dose prescription explicitely
  enforcing a ramp-like dose profile. The cranial and lower spinal part
  were done afterwards, taking into account the dose contribution of
  the upper spinal fields.

  \paragraph{Results} Using simple geometrical reasoning, we
  demonstrated that hot- and cold spots which arise from small
  displacement of one treatment field relative to the other treatment
  field can be reduced by taking two precautions: (a) widening the
  field overlap region, and (b) reducing the field gradient across the
  overlap region. The function with the smallest maximal gradient is a
  linear ramp. We present a treatment planning technique which
  yields the desired dose profile of the two contributing fields, and
  minimises dosimetric dependence on minor positional errors in
  patient set-up.

\end{abstract}
\maketitle

\section{Background and aim}

Craniospinal irradiation is a technically challenging task.  The
length of the planning target volume (PTV) exceeds the maximal size of
a treatment field, thus requiring some method of combining treatment
fields to treat the whole target.  While the standard set-up for
craniospinal irradiation is still based on the set-up described by Van
Dyk almost 40 years ago \cite{VanDyk:1977}, other modalities such as
intensity modulated radiation therapy (IMRT)
\cite{Parker:2007,Seppala:2010, Kusters:2011,Cao:2012,Studenski:2013,
  Wang:2013}, volumetric modulated arc therapy (VMAT)
\cite{Fogliata:2011,Chen:2012, Lee:2012,Myers:2013}, tomotherapy
\cite{Parker:2010,BandurskaLuque:2015}, and proton therapy
\cite{StClair:2004} are increasingly being examined as a possibility
for cranio-spinal irradiation.

The studies seem to agree that both IMRT and VMAT treatment planning
offer both a more conformal and a more homogeneous dose coverage of
the target and better sparing of some organs at risk (e.g., thyroid
gland) with respect to the traditional 3D~CRT approach, while at the
same time they raise concern about the increased dose to other organs
at risk, in particular lungs and kidneys.

A particular problem in the craniospinal irradiation are the field
junctions. Set-up inaccuracies of a few milimeters in the
cranio-caudal direction can result in large over- or underdosing. In
conventional set-up, moving the treatment field junction after a given
dose, usually every 9~Gy, has been adopted both for reducing dose
inhomogeneity and to minimise over- or underdosing which can occur due
to systematic errors; the technique is known as ``feathering''
\cite{Kiltie:2000}. Studies using IMRT employ a variety of
field-junction techniques: ``hybrid'' junction \cite{Parker:2007},
``jagged-junction'' \cite{Kusters:2011,Cao:2012}, and field overlap
\cite{Wang:2013} techniques, while studies using VMAT use almost
exclusively the field overlap technique \cite{Fogliata:2011,Lee:2012},
relying on the optimiser algorithm to arrive at a smooth field overlap
junction.

In the present study, we focus exclusively on the field junction
area. We show that a wider field junction can result in a smaller
dosimetric impact of a given positional error. We also demonstrate
that, left to itself, the dose optimiser algorithm may not arrive at a
dose profile which is the most robust towards small positional errors
in patient set-up. Finally, we present a treatment planning procedure
which reduces the dosimetric impact of positional errors.

\section{Materials and methods}

\subsection{Idealised field junction}

We start by showing that in two overlapping treatment fields, a linear
ramp is the dose profile which yields hot- and cold spots of the
smallest magnitude when one field is displaced by a small amount with
respect to the other. Denoting dose contributions of the two treatment
fields by $f(x)$ and $g(x)$, where $f(x) + g(x) = 1$, with $f(x) = 0$
and $g(x) = 1$ for $x \le 0$ and $f(x) = 1$ and $g(x) = 0$ for $x \ge
L$, we are interested in the deviation of the sum of both
contributions from unity, with one field being displaced by $\Delta
x$:
\[ 1 - \left( f(x) + g(x + \Delta x) \right) = f(x + \Delta x) - f(x) .
\]
When the displacement $\Delta x$ is small, we may expand the above
difference in a Taylor series and, retaining only the linear term in
$\Delta x$, we obtain:
\[ 1 - \left( f(x) + g(x + \Delta x) \right) \approx f'(x)\,\Delta x .
\]
Thus, at a given shift $\Delta x$, dose deviation is proportional to
the dose gradient $f'(x)$. Of all the functions raising from 0 to 1
over the distance $L$, the linear function has the smallest maximal
gradient.

Another quantity of interest would be the average deviation of the
dose from unity across the field junction:
\begin{align*}
  \frac{1}{L }\int_0^{L-\Delta x} \left[ 1 - \left( f(x) + g(x + \Delta x) 
    \right) \right] \mathrm{d}x &\approx 
  \frac{\Delta x}{L} \int_0^L f'(x) \,\mathrm{d}x \\
  &= \frac{\Delta x}{L} .
\end{align*}
Thus, for small displacements, when $\Delta x \ll L$ holds and terms
in $\Delta x/L$ higher than linear can be neglected, the shape of dose
profile only affects the magnitude of hot- and cold spots, but not the
average dose deviation.

\begin{figure}
  \centering\includegraphics[width=0.7\linewidth]{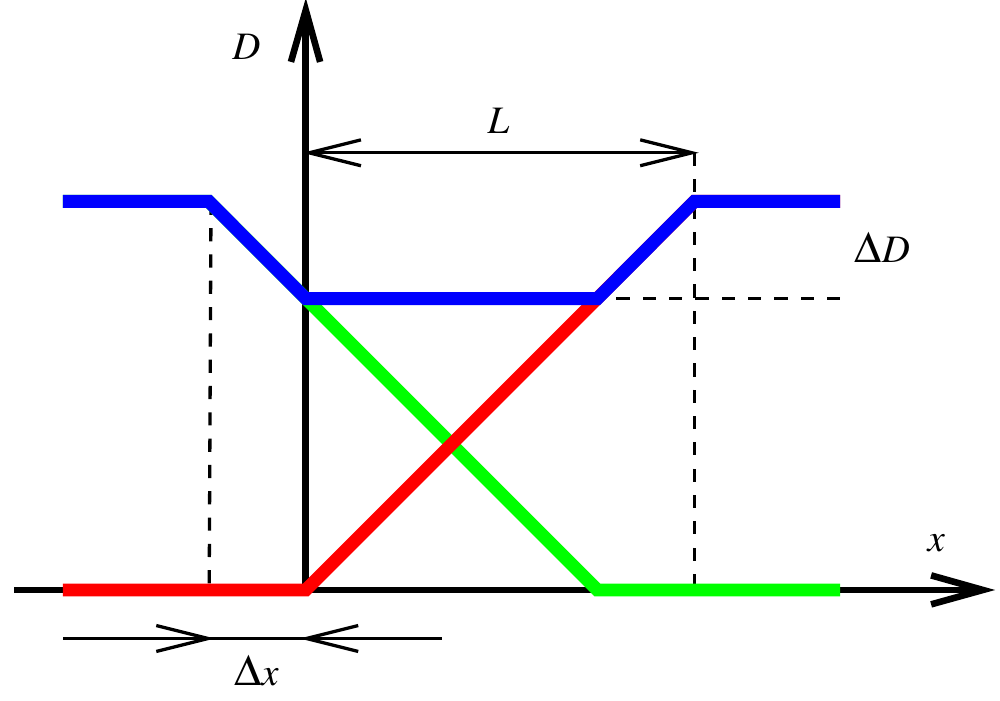}
  \caption{Dose profile across an idealised field junction in which
    the individual field contributions of are shown in red and green,
    respectively, and the total dose shown in blue.  A positional
    shift $\Delta x$ of one dose contribution with respect to the
    other results in a dose difference of $\Delta D$.}
  \label{fig:field-junction-delta}
\end{figure}

Fig.~\ref{fig:field-junction-delta} shows the dose profile across a
field junction in which $f(x)$ and $g(x)$ are represented by linear
ramps, one of them shifted with respect to the other. The
contributions of the two overlapping groups of fields (with VMAT, a
group of fields is usually a pair of arc fields; with IMRT it is
usually 5--7 fields) -- $f(x)$ and $g(x)$ -- are shown in red and
green, respectively, and the total dose shown in blue. In this
simplified example, we arrive at the same expression for dose
deviation $\Delta D$:
\[ \Delta D = 100\% \cdot \frac{\Delta x}{L} .
\]
As an illustration, a 5~mm positional shift is expected to result in a
5\% dose difference across a 10~cm field junction. The ratio $100\%/L$
is the dose gradient.

\subsection{Patient and target selection}

An adolescent female patient (18~yo) was selected as a test case. The
patient had undergone CT simulation in supine position with the head
and shoulders immobilised in a thermoplastic mask and the arms resting
comfortably at the patient's sides.  PTV encompassed the whole brain
and spinal cord down to S2 vertebra, resulting in a total length of
75.9~cm. The prescribed dose to this target was 30.6~Gy in 17
fractions.

\subsection{Treatment planning}

Treatment planning was done on Eclipse 10.0 treatment planning system,
with Varian Unique Performance equipped with Millenium MLC 120 and
Exact IGRT couch as the target treatment machine (Varian Medical
Systems, Palo Alto, CA, USA). The treatment plans were generated using
Progressive Resolution Optimiser algorithm PRO3
\cite{VarianEclipse:2009}, and the dose was computed with the
Analytical Aniso\-tropic algorithm (AAA) using 2.5~mm calculation grid
resolution.

To cover the whole PTV, 3 isocenters in the cranial, upper spinal
(Th3) and lower spinal (L2) region were chosen, with 23~cm separating
the cranial and the upper spinal isocenters, and 26~cm separating both
spinal isocenters.  To simplify patient positioning, the coordinates
of the three isocenters only differed in the cranio-caudal direction.
The field set-up used by center C in \cite{Fogliata:2011} was adopted,
with two partial arcs used at each isocenter, avoiding irradiation
from the anterior position by omitting the sector
$310^\circ$--$50^\circ$, and the collimator rotated to $10^\circ$ and
$350^\circ$, respectively. 

\begin{figure*}
  \centering\includegraphics[width=0.75\textwidth]{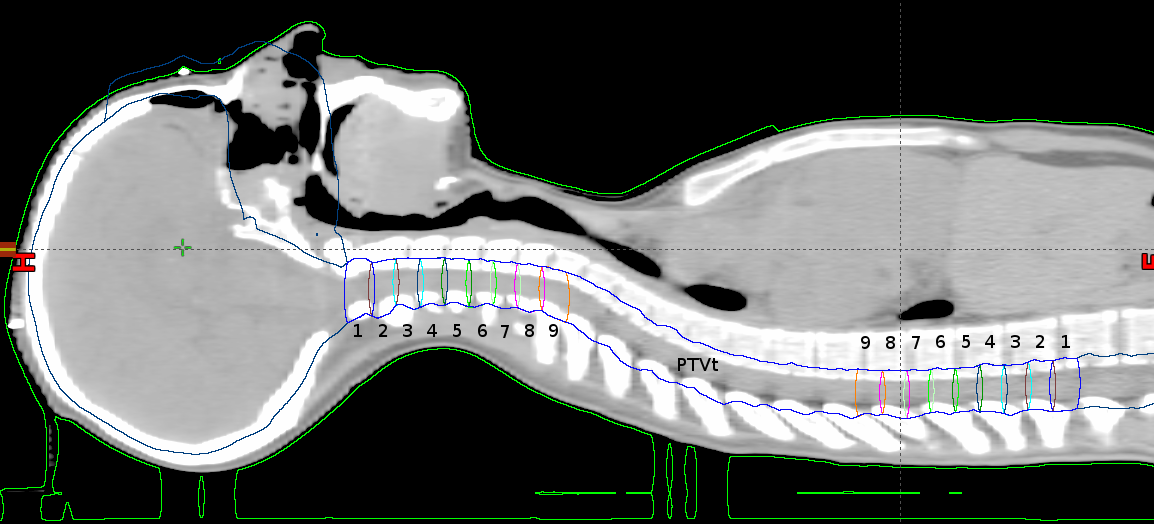}
  \caption{Sagittal view showing the auxilliary structures used for
    treatment planning.}
  \label{fig:aux-struct}
\end{figure*}

Dose optimisation was performed in two steps.  The treatment plan for
the upper spinal region was made in the first step.  Within the upper
spinal region, a central region (PTVt; 14.4~cm) was defined with a
homogeneous dose prescription, surrounded by two 10.8~cm transitional
regions (Fig.~\ref{fig:aux-struct}). Each of the transitional regions
was further divided into 9 subregions, each extending 1.2~cm in the
cranio-caudal direction.  The dose prescription in each subregion
gradually increased from the periphery towards the center; i.e. to
both outermost subregions (labelled as 1 in Fig.~\ref{fig:aux-struct})
a dose of 0--3.4~Gy was prescribed, to the adjecent regions (2) a dose
of 3.4--6.8~Gy was prescribed, and so on, up to the innermost subregions
(9), to which a dose of 27.2--30.6~Gy was prescribed.

After the treatment plan for the upper spinal region had been
optimised, the treatment plans covering the cranial and lower spinal
regions were made, taking into account the dose distrubution for the
upper spinal region, and filling up the dose to the prescribed value.

\subsection{Positional error simulation}

A positional error in the cranio-caudal direction was simulated by
taking an already optimised plan and making two modifications to it:
the isocenter position for the cranial pair of treatment fields was
moved 5~mm in the cranial direction, and the isocenter position for
the lower spinal pair of fields was moved 5~mm in the caudal
direction. The dose was recalculated while keeping the same monitor
unit count. In a complementary simulation, the isocenter position for
the cranial pair of treatment fields was moved 5~mm in the caudal
direction, and the isocenter position for the lower spinal pair of
fields was moved 5~mm in the cranial direction. The same procedure was
repeated for the positional shifts of $\pm 1$, $\pm 3$, $\pm 7$, and
$\pm 10$~mm.

\subsection{Treatment plan verification}

The treatment plan was verified using film dosimetry (Gafchromic EBT3;
Ashland, Wayne, NJ, USA).  A radiochromic film was mounted into the
slot for film dosimetry in the IBA MultiCube phantom (IBA Dosimetry,
Schwarzenbruck, Germany), with the phantom turned onto its side so
that the film lied in the sagittal plane.  Irradiated films were later
analysed using a web application for radiochromic film dosimetry
(Radiochromic.com, v2.2; \url{https://radiochromic.com/})
\cite{Mendez:2014}, using a correction for scanner response
variability \cite{Lewis:2015b}.

\section{Results}

\subsection{Dose profiles}

\begin{figure*}
  \centering\includegraphics[width=0.8\textwidth]{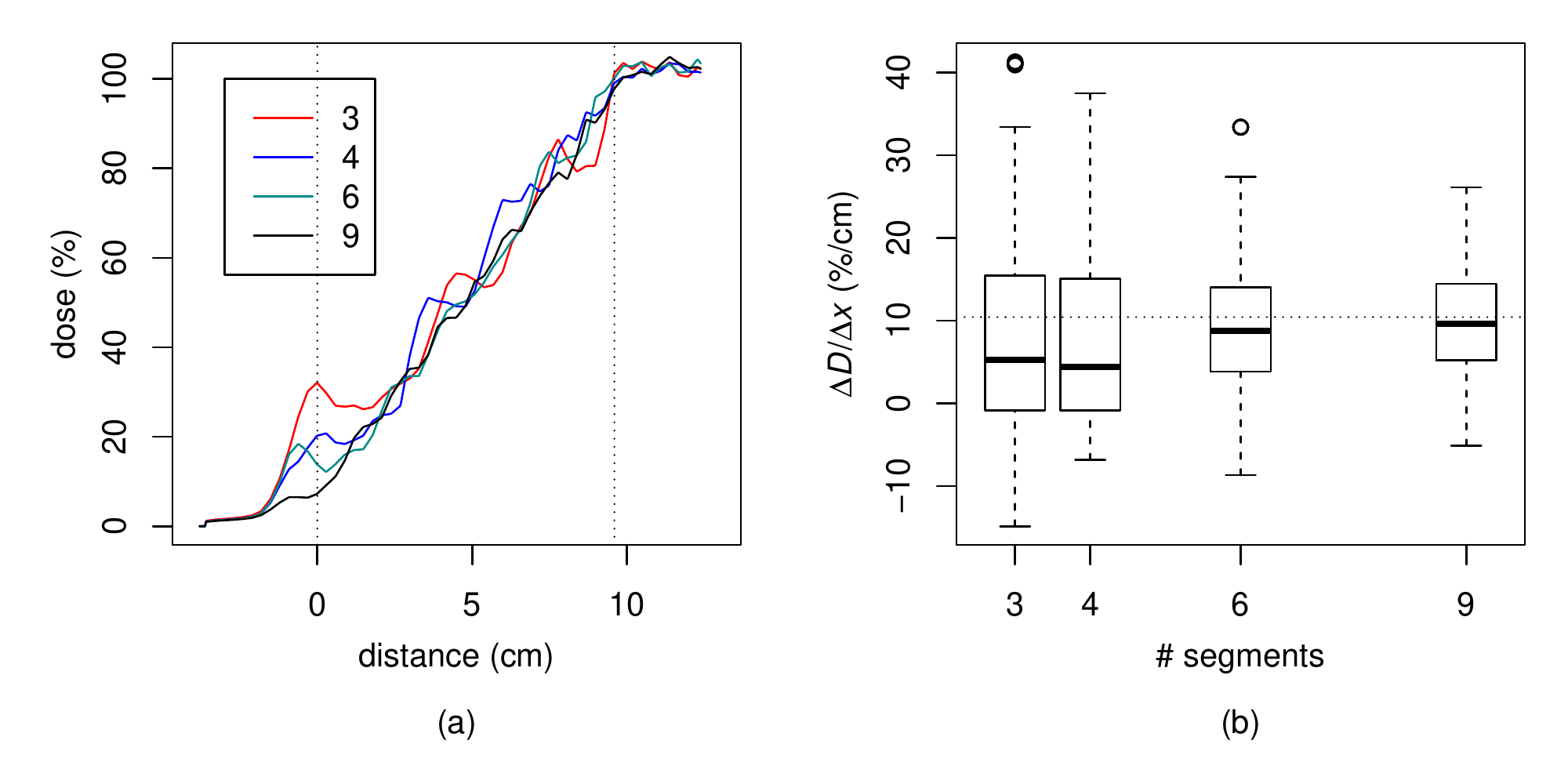}
  \caption{The calculated dose profile in the transitional region
    after optimisation in the case of dividing the transitional region
    into 3, 4, 6, or 9 segments (a). Distribution of dose gradient
    within the transitional region in the case of dividing the
    transitional region into 3, 4, 6, or 9 segments. The horizontal
    line corresponds to a uniform gradient (b).}
  \label{fig:ramp-compare}
\end{figure*}

In order to determine the minimum number of segments needed to obtain
a smooth linear ramp dose profile in the transition region, we ran a
separate experiment. In the dose optimisation step for the treatment
field for the upper spinal region, the transitional region (17 CT
slices, separated 6~mm apart, total length 96~mm) was divided into
either 3, 4, 6, or 9 segments. Figure~\ref{fig:ramp-compare}a shows
the dose profile contributed by the upper spinal fields to the
spinal-spinal junction.  Figure~\ref{fig:ramp-compare}b shows the
distribution of dose gradient along the transitional region for the
above mentioned segmentations, presented as box plots.  A linear ramp
has a uniform slope of 100\%/96~mm, or 10.4\%/cm. One can see that
increasing the number of segments leads to a dose profile closer to
the desired linear ramp.

\begin{figure*}
  \centering\includegraphics[width=0.85\textwidth]{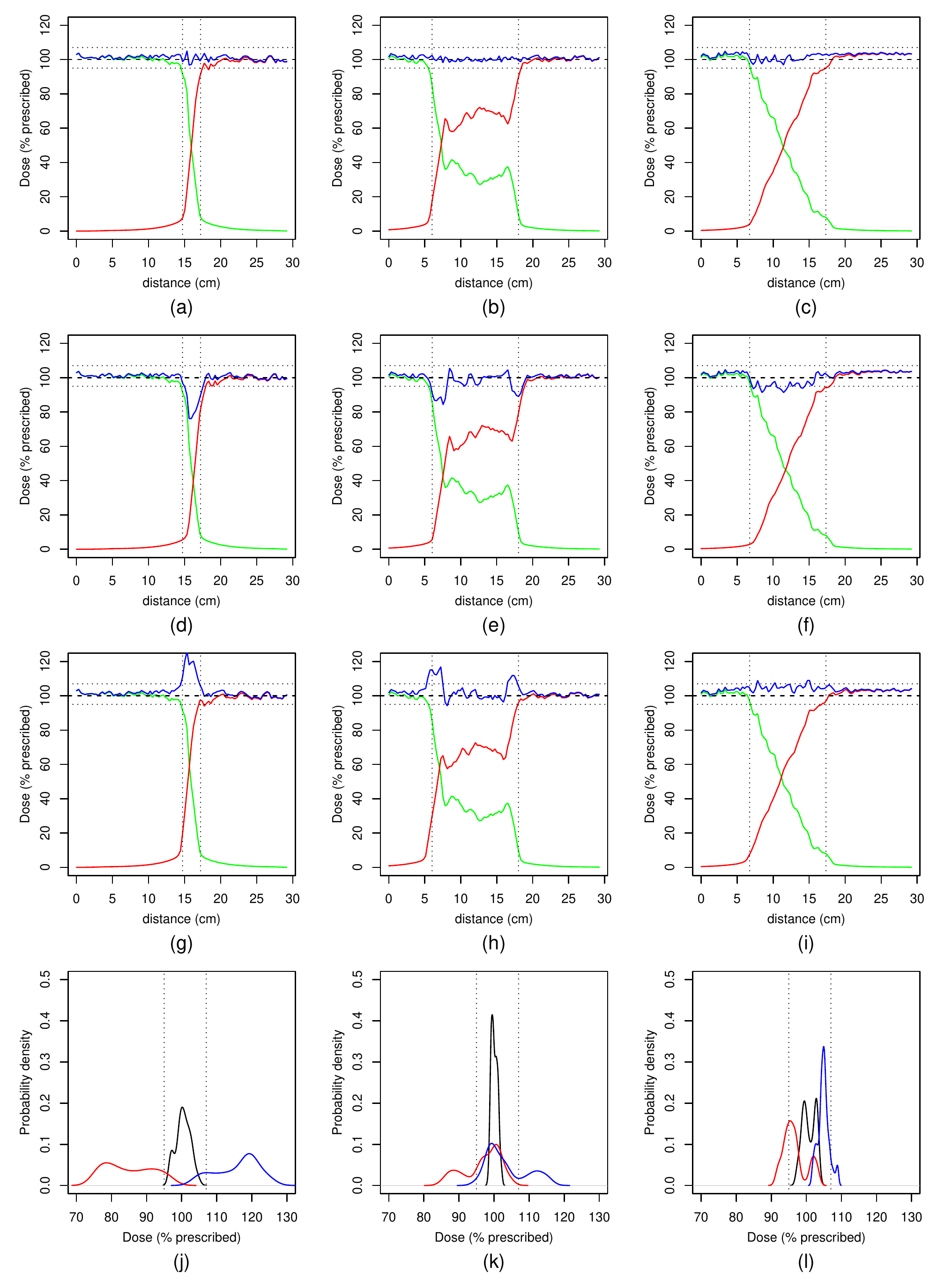}
  \caption{Dose profile across the field junction region in the
    spinal-spinal junction for a narrow junction (a), wide junction
    (b) and linear ramp junction (c). The dose contribution of the
    upper spinal field is shown as green, and the contribution of the
    lower spinal field as red. Figures (d--f) show the simulated
    effect of a $-5$~mm shift (apart in the cranio-caudal direction),
    and figures (g--i) show the effect of a $+5$~mm shift (towards
    each other). Figures (j--l) show the dose distribution in the
    field junction region; no positional shift (black), $-5$~mm shift
    (red), $+5$~mm shift (blue). Vertical lines in (a--i) show the
    extent of the field overlap region. Horizontal lines in (a--i) and
    vertical lines in (j--l) denote 95\% and 107\% of the prescribed
    dose, respectively.}
  \label{fig:torak-abdom-stik}
\end{figure*}

For different treatment plans, Figs.~\ref{fig:torak-abdom-stik}a--l
show a dose profile along a line between the same two points in the
patient's body, encompassing the spinal-spinal junction. The
contribution of the two upper spinal arc fields (clockwise and
counter-clockwise summed up) is shown in green, the contribution of
the two lower spinal fields is shown in red, and the total dose in
blue. Dose profiles in Figs.~\ref{fig:torak-abdom-stik}a,b have both
been obtained by using the progressive resolution optimiser PRO3 on
two overlapping pairs of VMAT fields, the only difference being that
in Fig.~\ref{fig:torak-abdom-stik}a, the overlap region is narrow
(2.5~cm), while in Fig.~\ref{fig:torak-abdom-stik}b, it is wide
(12.0~cm). Fig.~\ref{fig:torak-abdom-stik}c shows the dose profile
along a linear ramp junction (10.4~cm) described in the previous
section.

By comparing Fig.~\ref{fig:torak-abdom-stik}a and
Fig.~\ref{fig:torak-abdom-stik}b, it is clear that simply widening the
junction does not result in a significantly smaller dose gradient of
either upper or lower spinal field contribution.  Instead, the dose
optimiser algorithm produces a fairly constant total dose across the
junction using non-monotonous upper or lower spinal field
contributions, each of them still having areas of steep dose
gradient. Fig.~\ref{fig:torak-abdom-stik}c, however, shows that the
treatment planning technique presented in the previous section results
in a dose profile which is close to the ideal
(Fig.~\ref{fig:field-junction-delta}).

Figs.~\ref{fig:torak-abdom-stik}d--f show the behaviour of the field
junctions shown in Figs.~\ref{fig:torak-abdom-stik}a--c in a simulated
case in which the distance between the isocenters of the upper and
lower spinal fields has been increased by 5~mm. As expected, a large
drop in the total dose is observed in the regions where the upper and
lower spinal fields have a high dose gradient.  In a similar manner,
Figs.~\ref{fig:torak-abdom-stik}g--i show the behaviour of the field
junctions shown in Figs.~\ref{fig:torak-abdom-stik}a--c in a simulated
case in which the distance between the isocenters of the upper and
lower spinal fields has been decreased by 5~mm. Again, a large rise in
the total dose is observed in the regions of high dose gradient.

Figs.~\ref{fig:torak-abdom-stik}j--l show the dose distribution along
the dose profile line in the junction area for the field junctions
shown in Figs.~\ref{fig:torak-abdom-stik}a--c. Black lines correspond
to no positional shift, red lines to the distance between the
isocenters increased by 5~mm, and blue lines to the distance between
the isocenters decreased by 5~mm. While a 5~mm shift of one of the
fields with respect to the other can result in a $\pm 30\%$ change in
dose in the case of a narrow junction, this value drops to $\pm 15\%$
in the case of a wide junction, and down to $\pm 10\%$ in the case of
a linear ramp junction. Note this coincides fairly well with the
previous estimates for an idealised field junction: as the original
treatment plan aimed to cover the target with 95--105\% of the
prescribed dose, a 5~mm positional shift over a 10~cm junction is
expected to pull the dose up or down by another 5\%, depending on the
direction of the shift.

Counting merely the percentage of points staying within the
$[95\%,107\%]$ dose range after a $\pm5$~mm shift is performed, the
linear ramp dose profile does not offer a significant improvement over
other dose profiles at comparable junction widths. A narrow junction
clearly shows the worst results, with only 10\% and 13\% points
staying within the prescribed dose range after the the distance
between the isocenters increased or decreased by 5~mm,
respectively. With a wide junction, these values climb up to 73\% and
71\%, respectively, while in the case of a linear ramp junction, they
are 68\% and 91\%. Thus, while overall a linear ramp junction performs
slightly better, we don't consider this difference as important as the
magnitude by which the dose departs from the prescribed range.

\begin{figure}
  \centering\includegraphics[width=\linewidth]{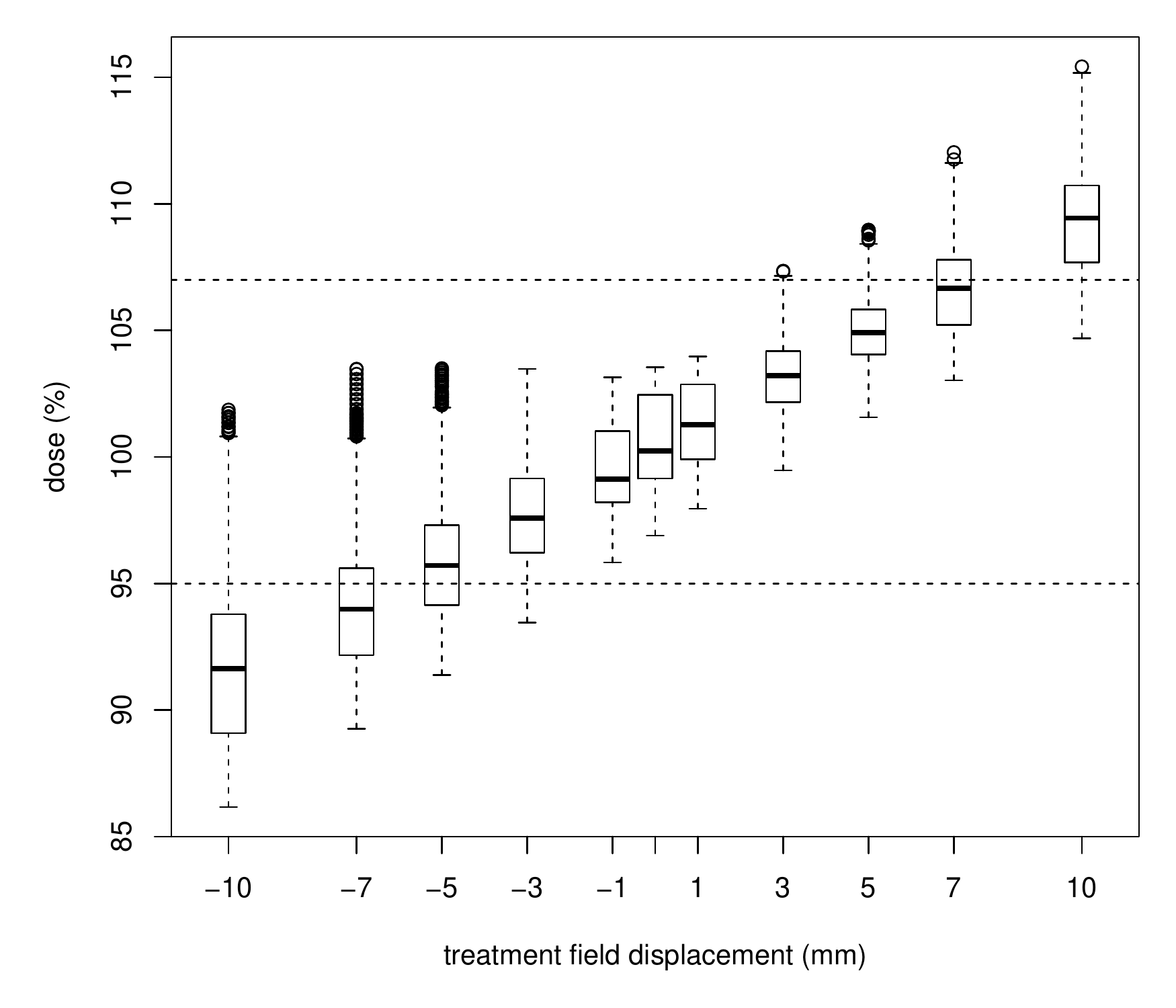}
  \caption{Dose distribution along the dose profile within the
    transitional region of the spinal-spinal junction for different
    values of positional shift of the lower spinal treatment field
    pair. Horizontal lines denote 95\% and 107\% of the prescribed
    dose, respectively.}
  \label{fig:displac-dep}
\end{figure}

For a linear ramp junction, the sensitivity of the dose coverage
within the transitional area to the magnitude of positional shift has
been assessed (Figure~\ref{fig:displac-dep}). In this experiment,
after dose optimisation, the isocenter of the lower spinal treatment
field pair was shifted $\pm 1$ to $\pm 10$~mm in the cranial or caudal
direction, and the dose was recalculated. Dose distribution within the
transitional area of the spinal-spinal junction is shown as a box
plot. One can see that a positional shift up to $\pm 3$~mm generally
leads to dose distributions which are within the $[95\%, 107\%]$ range,
which is considered acceptable \cite{ICRU:Report50}, while larger
positional shifts, even though they may still yield an acceptable dose
distribution in some circumstances, are generally considered
unacceptable.

\subsection{Dosimetric influence of arm movement}
Using the whole arc except the sector $[300^\circ,50^\circ]$ for
irradiation means irradiating through the patient's arms. While this
has been done with suitable immobilisation \cite{Mancosu:2012}, it is
generally not advised, because of the dosimetric inaccuracies induced
by a non-deliberate change in the patient's arms position. In a
separate simulated experiment, we estimated the effect of a complete
omission of the patient's arms: after optimising the treatment plan,
the same treatment plan was recalculated on another structure set,
identical to the original one with the exception of the patient's arms
being excluded from the calculation volume. As expected this produced
a slightly higher dose at the same MU value, with $D_{50\%}$ for the
spinal PTV rising from 101.3\% of the prescribed dose to 103.1\% of
the prescribed dose. Similarly, $D_{98\%}$ rose from 94.8\% to 96.8\%
of the prescribed dose, and $D_{2\%}$ rose from 104.2\% to 107.5\% of
the prescribed dose. This is the extreme case; we can expect that a
minor change in position would induce a dosimetric error smaller than
this worst-case scenario. The relatively low impact of arm
displacement is consistent with a relatively low dose received by the
patient's arms; even though no special restrictions to the dose in the
arms were used, the maximal dose to both humeri stayed at around 10\%
of the dose prescribed to PTV.

\subsection{Treatment plan verification}

\begin{figure}
  \centering\includegraphics[width=9cm]{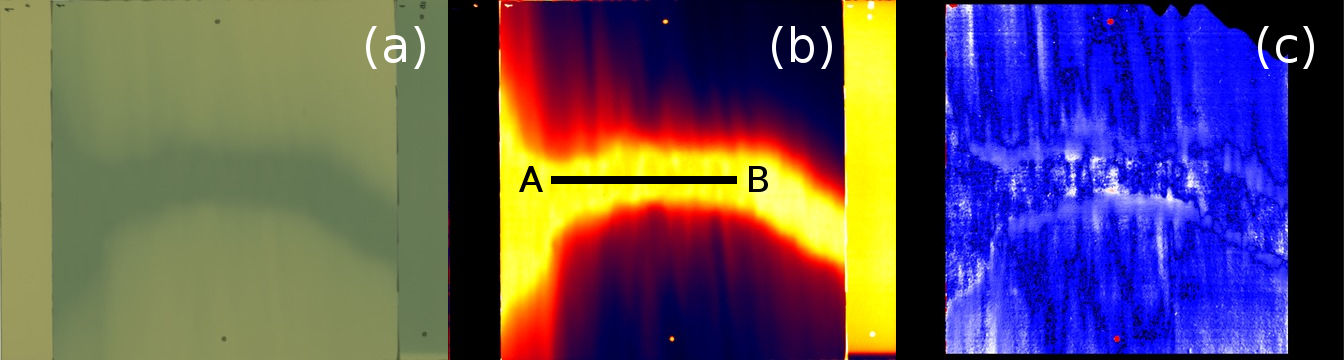}
  \centering\includegraphics[width=9cm]{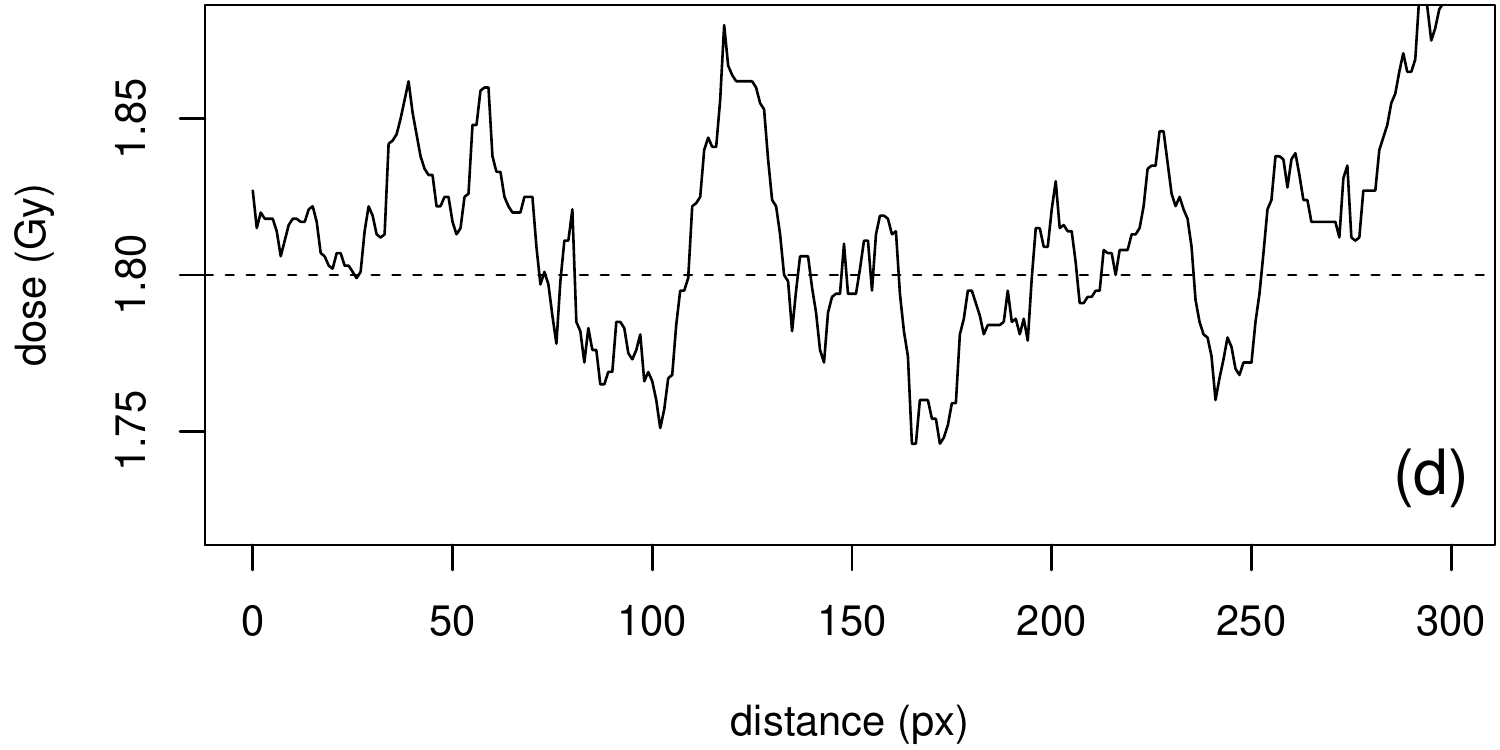}
  \caption{QA verfification of the cranio-spinal field junction; (a)
    irradiated radiochromic film with two vertical stripes containing
    a non-irradiated film and a film exposed to 1.8~Gy, (b) dose
    calculated from the radiochromic film readout, (c) gamma index
    analysis of the match between the dose obtained with the
    radiochromic film and the dose plane exported from the treatment
    planning system, (d) dose profile along the line A--B.}
  \label{fig:multicube-film}
\end{figure}

Fig.~\ref{fig:multicube-film} shows the results of treatment plan
verification using film dosimetry; the irradiated film scan
(Fig.~\ref{fig:multicube-film}a), the calculated dose
(Fig.~\ref{fig:multicube-film}b), the gamma index analysis
(Fig.~\ref{fig:multicube-film}c) of the match between the dose
obtained with the radiochromic film and the dose plane exported from
the treatment planning system, and the dose profile across the field
overlap region (Fig.~\ref{fig:multicube-film}d).  For gamma index
analysis \cite{Low:1998}, 3\% dose tolerance and 3~mm positional
tolerance was used along with a threshold value set at 10\% of
$D_\mathrm{max}$ and global normalization, resulting in the average
$\gamma$ value of 0.16 and 99.7\% points passing $\gamma < 1$
criterion. Points with $\gamma < 1$ are shown in blue, and points with
$\gamma > 1$ are shown in red.

\section{Discussion}

\subsection{Prone vs. supine patient position}

While the original set-up for craniospinal irradiation
\cite{VanDyk:1977} employed a prone patient position, there are
numerous studies advocating the supine patient position
\cite{Hawkins:2001,Slampa:2001,Slampa:2006,Parker:2006}. After judging
the advantages and the disadvantages of both set-ups, we decided to
employ it as well.  The obvious advantage of the prone position --
easier visualisation of treatment field junctions -- became less
important with the advancement of image-guided radiotherapy, which
made the supine position a viable alternative, allowing for greater
patient comfort and easier access for the paediatric patients
requiring anaesthesia.

\subsection{Dose profile in field overlap area}

While the studies treating the craniospinal axis using IMRT often paid
special attention to achieving a desired dose profile in the field
junction area (e.g., staircase, \cite{Kusters:2011,Cao:2012}), the
studies using VMAT usually seem to rely on the dose optimiser to
obtain a suitable dose profile. As we have shown earlier, this often
results in non-monotonous dose profiles.

However, a recent study \cite{Mancosu:2015b} dealing with the
robustness of treatment plans with respect to positional shifts of the
patient claims that the authors specifically aimed for a sigmoidal
dose profile in the field overlap region. In another study on
localization errors \cite{Myers:2013} the authors propose a staircase
dose profile in order to minimize such errors. An interesting feature
of this study is that the authors actually arrived at a linear ramp
profile when the junction region was too narrow to support staircase
dose profile (\cite{Myers:2013}, Fig.~4a), but brushed it off as a
mere curiosity without realising that it gives even better results
than the technique they propose. Using a linear ramp instead of a
staircase dose profile, we predict the $\pm 10\%$ spikes observed by
Myers et al. (\cite{Myers:2013}, Fig.~3b,d), which correspond to a
$\pm 5$~mm shift applied to a 116~mm overlap region, would drop down
to $\sim 5\%$.

\subsection{Maximal vs. average positional errors}

Earlier studies \cite{Mancosu:2013,Mancosu:2015b} have proven that the
position shifts in the cranio-caudal direction result in greater
dosimetric changes than the shifts in the other two
directions. Consequently, we limited our study to the cranio-caudal
direction. A comparision of the dosimetric changes reported in
\cite{Mancosu:2013,Mancosu:2015b} to the ones reported here is
difficult for two reasons: (a) we examined local effects (hot- and
cold spots) rather than their influence on the volume-dose histogram,
and (b) we evaluated the dosimetric impact of the maximal shift, i.e.,
$\pm 5$~mm, while the authors in \cite{Mancosu:2013,Mancosu:2015b}
chose three random values from the
$[-5\;\textrm{mm}$, $+5\;\textrm{mm}]$ range and used their mean and
standard deviation. Assuming that the samples are distributed
uniformly, the mean of several uniform distribution is a rectangular
mean distribution, sometimes called Bates distribution. For 3
uniformly distributed samples taken from the
$[-5\;\textrm{mm},+5\;\textrm{mm}]$, the expected mean and standard
deviation is $0\pm 1.67\;\textrm{mm}$. It can be argued that this
approach better reflects the reality of a clinical department; however
we have deliberately chosen more direct observables which allow for
easier analysis.  A 21\% increase of the maximum dose reported in
\cite{Mancosu:2015b} is however close to the expected value of 25\%
for a 5~mm shift in the cranio-caudal direction applied to a 20~mm
overlap region.

\subsection{Quality assurance of medical accelerators}

Craniospinal irradiation, in particular when performed in its
conventional set-up, is a technique which leaves little margin for
errors. As its requirements are close to the tolerance levels
specified in the quality assurance (QA) recommendations for medical
accelerators, it is important to consider the relevant parameters. The
conventional set-up relies on the jaw position for half-beam block and
the treatment couch position; tolerance levels for them are 1~mm and
2~mm, respectively, both are checked monthly \cite{Klein:2009}.

With image-guided techniques, assuring that the imaging system
isocenter coincides with the treatment isocenter is of prime
importance. The suggested tolerance levels here are 1.5--2~mm, checked
monthly \cite{Yoo:2006,Guan:2009}. The inaccuracies introduced by
mis-alignment result in systematic errors which do not cancel out by
day-to-day variations.

\section{Conclusions}

Using an idealised schematic field junction with partial field
overlap, we have shown that hot- and cold spots which arise from small
displacement of one treatment field relative to the other treatment
field can be reduced by taking two precautions: (a) widening the field
overlap region, and (b) reducing the field gradient across the overlap
region.  The function with the smallest maximal gradient is a linear
ramp, and we presented a treatment planning technique for craniospinal
irradiation which yields the desired dose profile of the two
contributing fields, and minimises dosimetric dependence on minor
positional errors in patient set-up.  The treatment planning technique
was developed using VMAT, but can be equally well applied to every
IMRT-derived technique.

%
%
%
\section*{Financial disclosure}

PP acknowledges the financial support from the Slovenian Research
Agency through research grants P3-0307 and P1-0389.

\section*{Acknowledgements}

The authors thank A.~\v{S}arvari for helpful discussions concerning
quality assurance.

\section*{References}
\bibliographystyle{elsarticle-num}

\end{document}